\documentclass{article}

\usepackage{arxiv}

\usepackage[utf8]{inputenc} 
\usepackage[T1]{fontenc}    
\usepackage{hyperref}       
\usepackage{url}            
\usepackage{booktabs}       
\usepackage{nicefrac}       
\usepackage{microtype}      
\usepackage{amsmath}
\usepackage{amssymb}
\usepackage{amsfonts}
\usepackage{cleveref}       
\usepackage{graphicx}
\usepackage{doi}
\usepackage[backend=biber, sorting=none]{biblatex}
\usepackage[british]{babel}
\usepackage{csquotes}
\usepackage{orcidlink}
\usepackage{subcaption}
\usepackage{float}

\emergencystretch=\maxdimen
\title{Evaluating the Socioeconomic Status of a Large Social Event Attendees}

\author{Kerecsen Szabó \orcidlink{0000-0002-6319-803X}, Gergő Pintér \orcidlink{0000-0003-4731-3816}, Imre Felde \orcidlink{0000-0003-4126-2480} \\
	Óbuda University, John von Neumann Faculty of Informatics, Budapest, Hungary \\
	\href{mailto:kerecsen.szabo@stud.uni-obuda.hu}{kerecsen.szabo@stud.uni-obuda.hu}, \href{mailto:pinter.gergo@nik.uni-obuda.hu}{pinter.gergo@nik.uni-obuda.hu}, \href{mailto:felde.imre@nik.uni-obuda.hu}{felde.imre@nik.uni-obuda.hu}
}

\begin{document}

\maketitle

\begin{abstract}
In this study, Call Detail Records (CDRs) from downtown Budapest were analysed, focusing on a large-scale event in August 2014. The attendees of the main event of the Hungarian State Foundation Day have been analysed based on their Socioeconomic Status (SES). This paper proposes an approach to estimating SES by the price and age of the subscribers' phones, obtained by fusing a mobile phone property database with the CDRs. We have found some tendencies between the attendees based on the location, from where they watched the fireworks. However, the results do not show significant differences in this geographical granularity.
\end{abstract}


\section{Introduction}

Since the mid-2000s, mobile telephones have become an inseparable part of our lives. The constant communication between a device and the mobile network leaves an involuntary trace of our activities. The spatio-temporal logs of our journeys and additional communication device information establish a promising model to evaluate human mobility and socioeconomic customs.

For the last two decades, uncovering underlying information from mobile phone records has been a developing research field. Data scientists, spatial data analysts, physicists and applied mathematicians pay more and more attention to discovering cellular data. Interpreting large amounts of Call Detail Records (CDRs) into useful information requires various tools and expertise. In the last ten years, dozens of research groups have published several major research journals discussing different applications of mobile network data analysis.

St. Stephen's Day is celebrated in Hungary every year on the 20$^{th}$ of August. Tens of thousands of people visit the capital for its all day long celebrations and the main event, a 30-minute-long firework show. The main area of the event includes three bridges and the embankments on the Danube for approximately three kilometres. With great views from the Buda and Pest embankments and the Castle District, these areas should show a significant spike in cellular activity and will be the primary subject of the analysis.

\section{Related works}
Using call detail records spanning over 52 weeks, accumulated over two-week-long periods, Gonzalez \textit{et al.} analysed 100,000 randomly selected individuals' movements over half a year in Europe. They introduced basic human mobility patterns and discovered that most individuals travel only short distances, and just a few move over hundreds of kilometres. The study approximated the probability density function of travel distances with a truncated power-law \cite{gonzalez2008understanding}.

Candia \textit{et al.} carried out a comprehensive study on the mean collective behaviour of individuals and examined how space and time deviations can be described using known percolation theory tools. They also proved that the inter-event time between consecutive calls is heavy-tailed, agreeing with previous studies on other human activities \cite{candia2008uncovering}.

A typical application of CDR processing is the sizeable social event detection and estimating the attendance during mass gatherings \cite{wirz2013probing, mamei2016estimating, barnett2016social, hiir2019impact}.

This study builds upon previous works at John von Neumann Faculty of Informatics, Óbuda University. The 2014 State Foundation Day data set has already been analysed \cite{pinter2018evaluation} regarding the large social event. Whereas in this work, the socioeconomic status of the attendees is studied.

Using mobile phone prices as Socioeconomic Status (SES) indicators has been proved to work well by Sultan \textit{et al.} in \cite{sultan2015mobile}. They identified areas in Pakistan where more expensive phones appear more often using indicators of accessibility to services, infrastructure, hygiene and communication. Their model performed with an absolute Pearson's correlation coefficient $> 0.35$ and p-value $< 0.01$.

In an earlier study, Pintér \textit{et al.} evaluated the connection between individuals' financial status and mobility customs. The authors used the radius of gyration, entropy, and Euclidean distance between home and work locations as mobility indicators and applied data fusion methods with average real estate prices to determine the influence of wealth on mobility customs \cite{pinter2021evaluating}.

Regarding socioeconomic status analysis, Pintér \textit{et al.} evaluated football fans' SES using their mobile phone details in Budapest during the 2016 UEFA European Football Championship. They eliminated CDRs from Subscriber Identity Modules (SIMs) during data preprocessing, which did not operate in mobile phones using Type Allocation Code (TAC) databases. \cite{pinter2021analyzing} In another work, they analysed subscribers' wake-up times and explained how it correlates with their socioeconomic status. The analysis demonstrated a strong positive connection between the two indicators. They also showed that the mobile phone prices in the TAC database might have depreciated \cite{pinter2022awakening}.

\section{Mobile network data}

A CDR data set usually contains a caller ID, the cell tower it is connected to, its location, and a timestamp. Additional information on the purpose of communication, device type, and SIM holders' details help investigate more than just trajectories and cell densities.

The data set used for this research was obtained by Vodafone Hungary Ltd. The number of active SIMs was 11,540,058 in Hungary, of which Vodafone had an estimated 22.7 per cent of the market share in June 2014 (Hungarian National Media and Infocommunications Authority). These CDRs contain anonymous logs of customers' calls and text messages in Budapest, Hungary and its suburban areas, over approximately 525.14 km$^2$ (Hungarian Central Statistical Office).

The data set was collected between the 18$^{th}$ and 22$^{nd}$ of August 2014. A total amount of 191,528,883 records have been logged, between 8,890 cells. Three comma-separated value files have been acquired for the analysis in this research. The first one contained call data records, the second and third are supplementary information about cells and devices.

The CDRs in this data set consist of a timestamp, a hashed device identification (ID), a hashed cell ID and a type allocation code. TAC is the initial eight-digit segment of a device's International Mobile Equipment Identity (IMEI), uniquely identifying a particular device. In this data set, CDRs are of active call record type, meaning a record was made when the user was making a call or sending/receiving a text message. Unfortunately, the data set does not contain information on cell switching, which would make more granulated data possible.

The supplementary cell lookup table contains cell IDs and positions as 2D coordinates in decimal degrees format. Cells at the same location were merged into base stations for further analysis, and the corresponding CDR's cell ID values were updated. Uniting particular cells is necessary for more straightforward data analysis. Nevertheless, cell tower antennas planted at an exact location might face different directions, but we do not have this information.

The device table contains a hashed device ID, the customer's age, gender, whether it is an individual or a business, and the subscription type (prepaid or postpaid). Some age and gender information is missing due to privacy restrictions.

\section{Methodology}

\begin{figure}[H]
	\centering
	\includegraphics[width=0.5\linewidth]{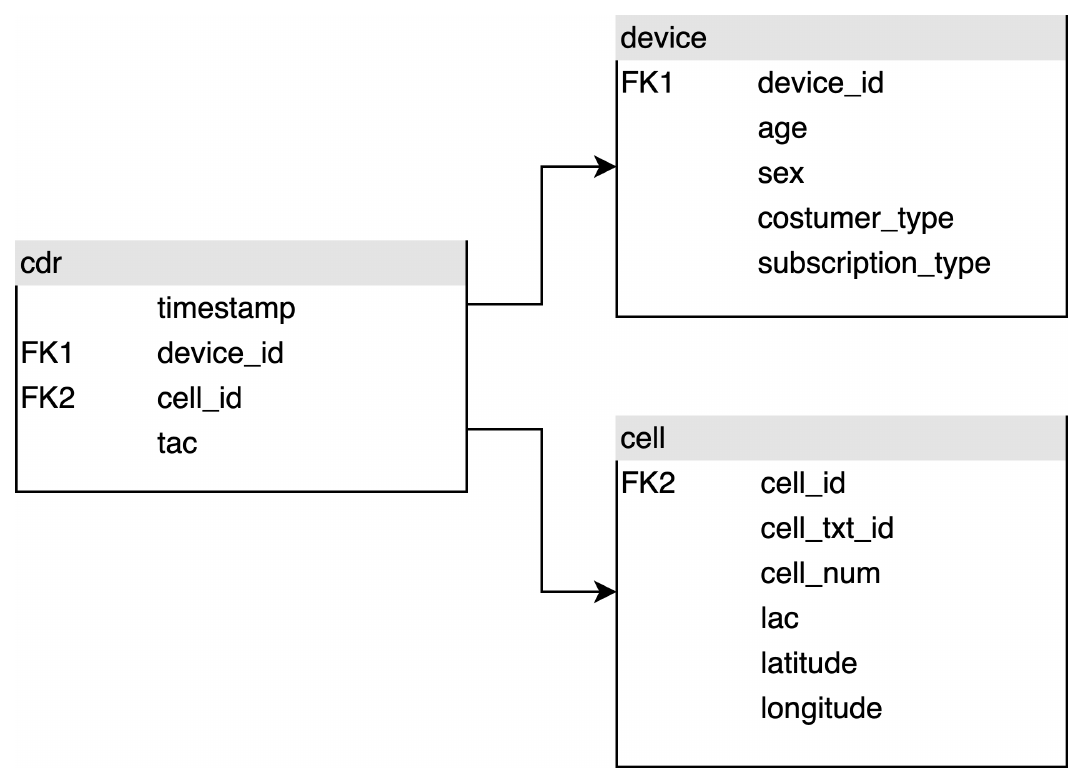}
	\caption{The database structure showing CDR, device and cell tables with foreign key connections.}
	\label{fig:database}
\end{figure}

During data cleansing, unnecessary spaces have been removed from the end of the lines. Furthermore, device and cell IDs were hashed information, a fundamental step for user privacy. On the contrary, it has a low information density and thus has been replaced with incrementing integer values. The reassigning does not affect the contained information, but is favourable to reducing data size. Due to the considerable amount of CDRs, the comma-separated value text files have been loaded into an SQLite database, with the scheme illustrated in Figure \ref{fig:database}. To speed up data acquisitions, indices were created on timestamps, IDs and TACs.

Although latitude and longitude values are in string format, there is no need for the original precision of 13 decimal points. This would mean $\mu m$ resolution, while anything above the sixth decimal place is useless in this application. The unnecessary information can be discarded to save space and increase query speeds.

As a socioeconomic status indicator, the analysis used relative mobile phone ages in months to the event (August 2014) and phone release prices in EUR. Information on resolving the TACs is from the data provided by 51Degrees fused \cite{pinter2021analyzing} with the GSMArena database \cite{gsmarena}.

\begin{figure}[H]
	\centering
	\includegraphics[width=0.65\linewidth]{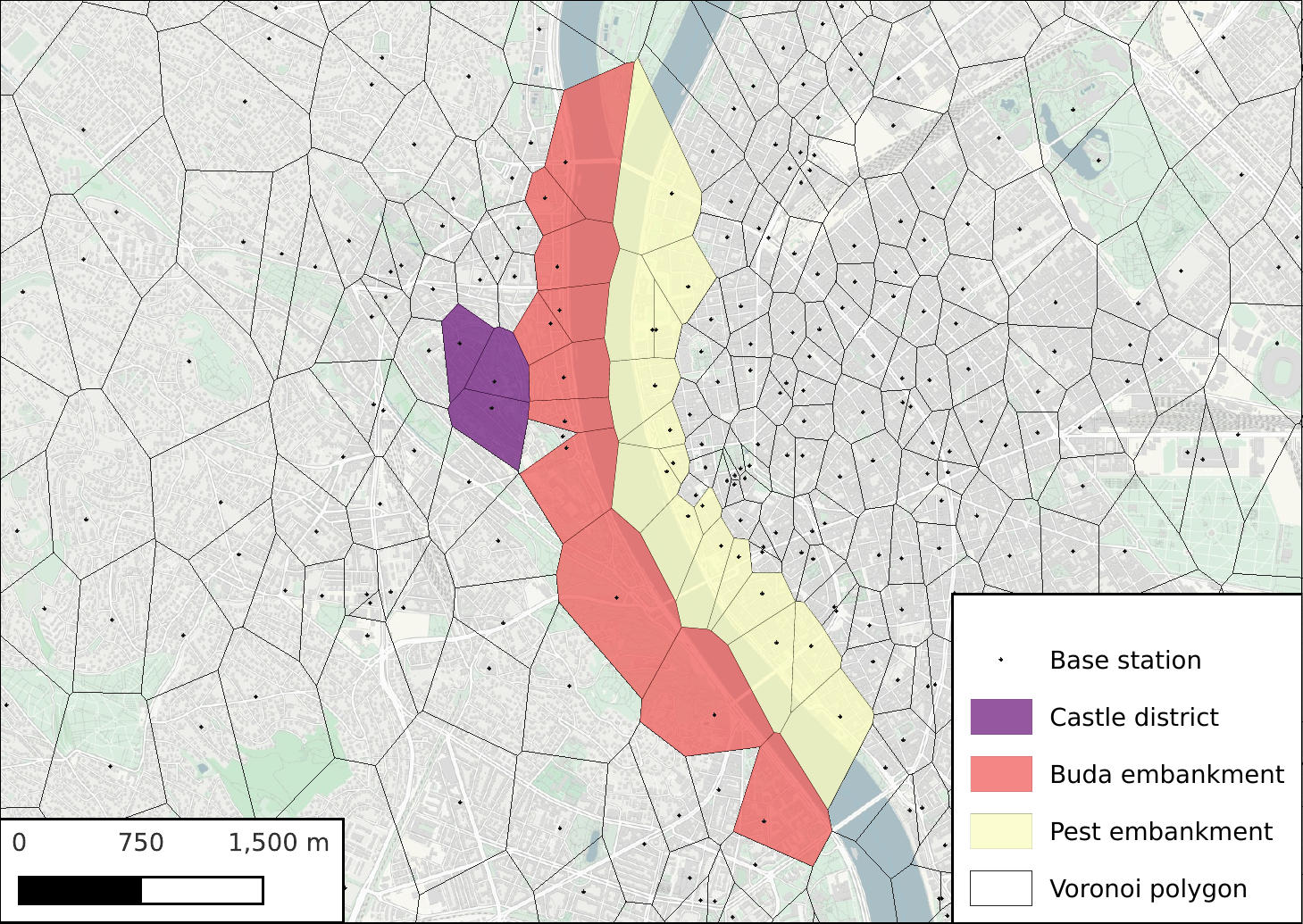}
	\caption{The selected cells for the large social event analysis.}
	\label{fig:cellsfirework}
\end{figure}

A data processing framework has been developed to acquire the relevant CDRs with the paired SES indicators. The first criteria are spatially being in a position that could indicate attendance at the fireworks. Cells along the two sides of the Danube are expected to be the primary cells servicing the attendees' mobile phones on the embankments. These cell IDs were selected manually due to the uneven separation line on the river. Any other cells that might support the selected areas are determined by a 250 m radius around the main event area, as visible in Figure \ref{fig:cellsfirework}. Four cells were removed from the evaluation due to the insignificant activity count (less than 500) during the event. Extracting the selected cell IDs helps filter and transform the large CDR database table into a more manageable format. The temporal filtration rule for the fireworks data will be $\pm$ 30 mins to the actual event. This will include possible additional users who did attend the venues but did not use their phones during the half-hour show.

User data from the device database table can be joined on the CDRs' device ID fields. This gives us the ability to analyse age and gender distributions in selected groups. The selected CDRs have been joined on the corresponding TAC values from the merged mobile phone property database for the SES indicator aggregation. For further analysis, relative ages of the appearing phones have been calculated from release dates and months to the event date (August 2014).

\begin{figure}[H]
	\centering
	\begin{subfigure}[t]{0.49\linewidth}
		\centering
		\includegraphics[width=\linewidth]{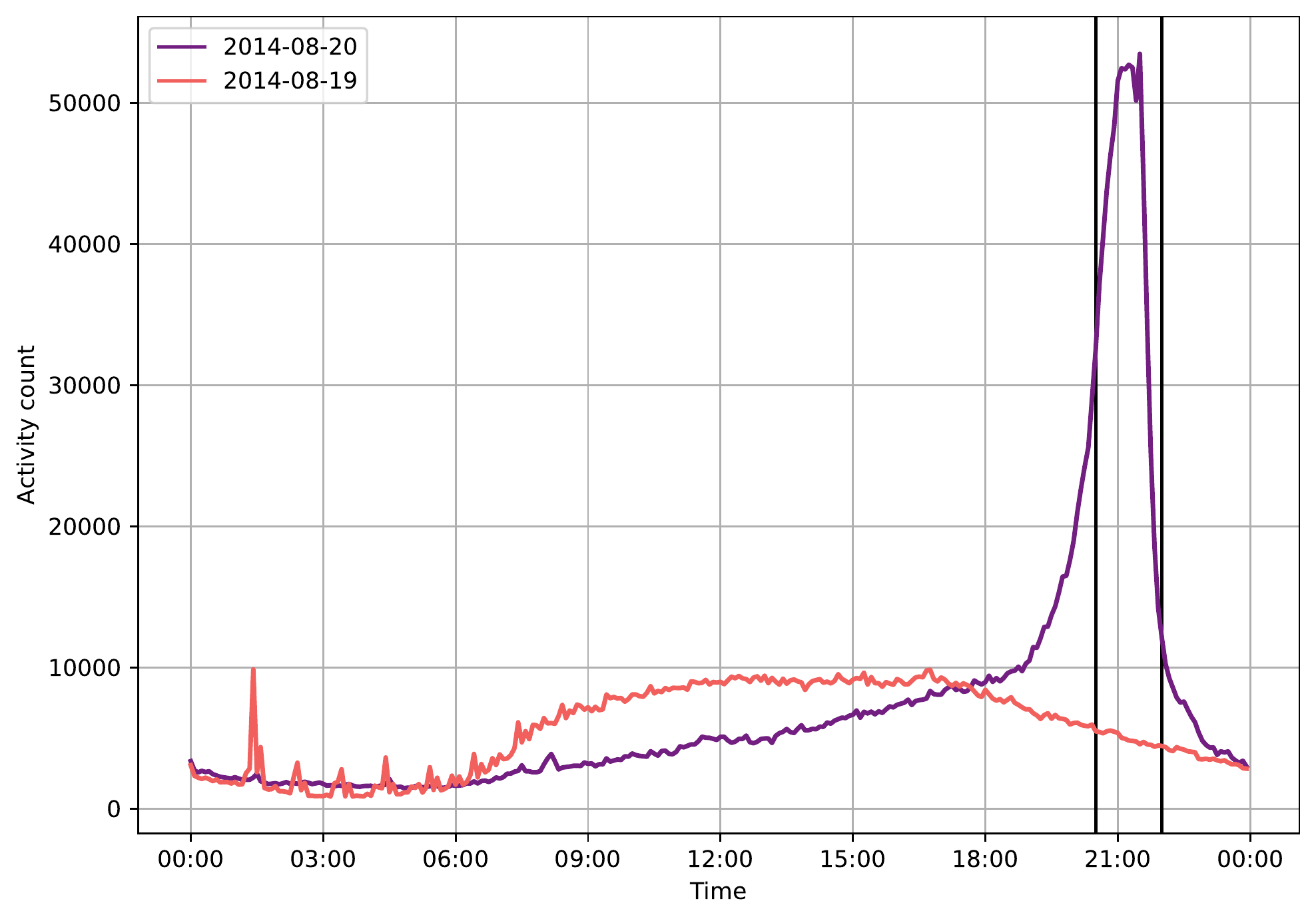}
		\caption{Daily cell activities in the studied area.}
		\label{subfig:event_cells}
	\end{subfigure}
	\hfill
	\begin{subfigure}[t]{0.49\linewidth}
		\centering
		\includegraphics[width=\linewidth]{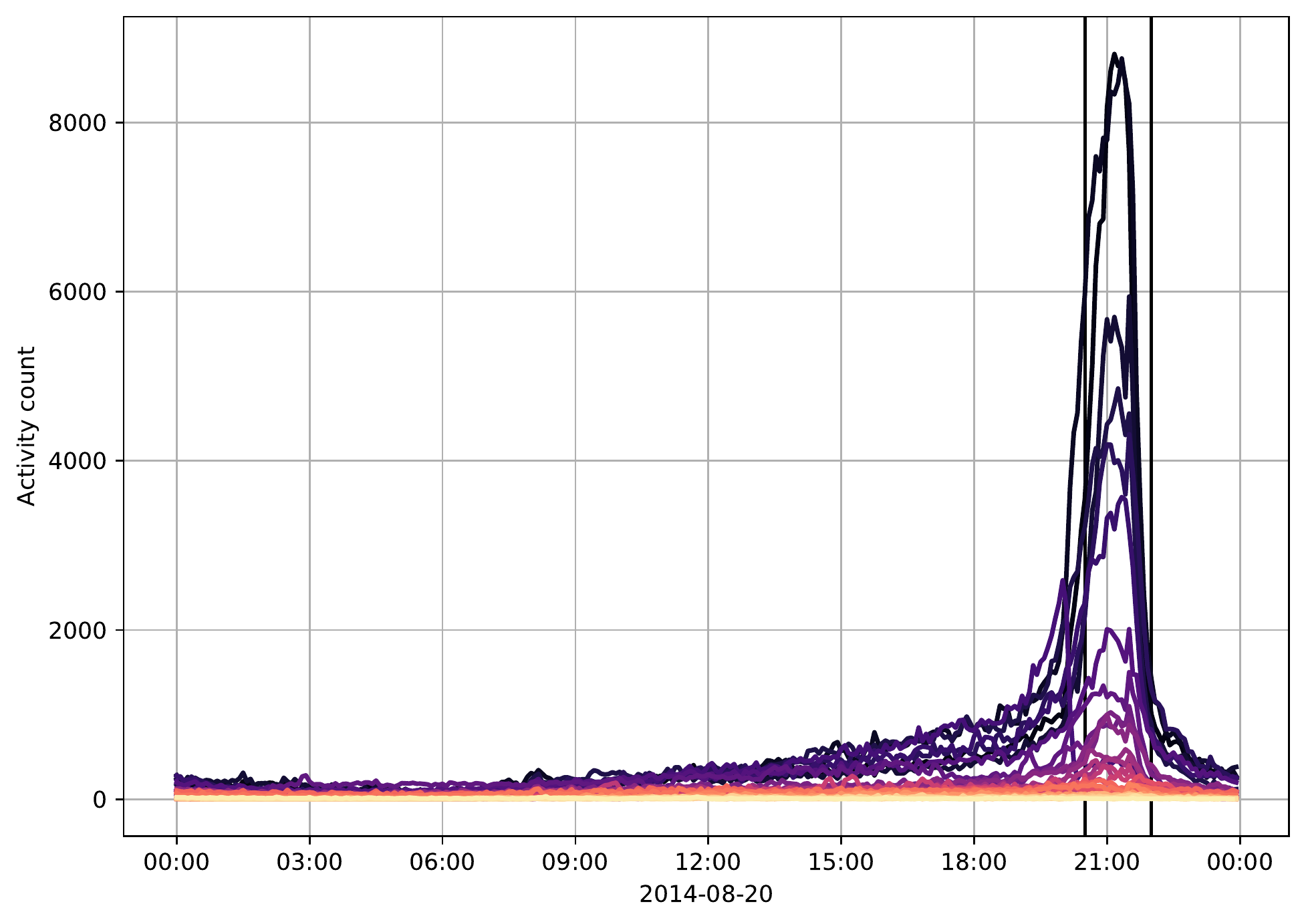}
		\caption{Daily cell activities between the studied cells.}
		\label{subfig:event_normal}
	\end{subfigure}
	\caption{Cell activities showing the extra mobile network load due to the State Foundation Day celebrations on the river Danube embankments and the Castle District in Budapest.}
	\label{fig:event_day}
\end{figure}

\section{Results}

\begin{figure}[H]
	\centering
	\begin{subfigure}[t]{0.49\linewidth}
		\centering
		\includegraphics[width=\linewidth]{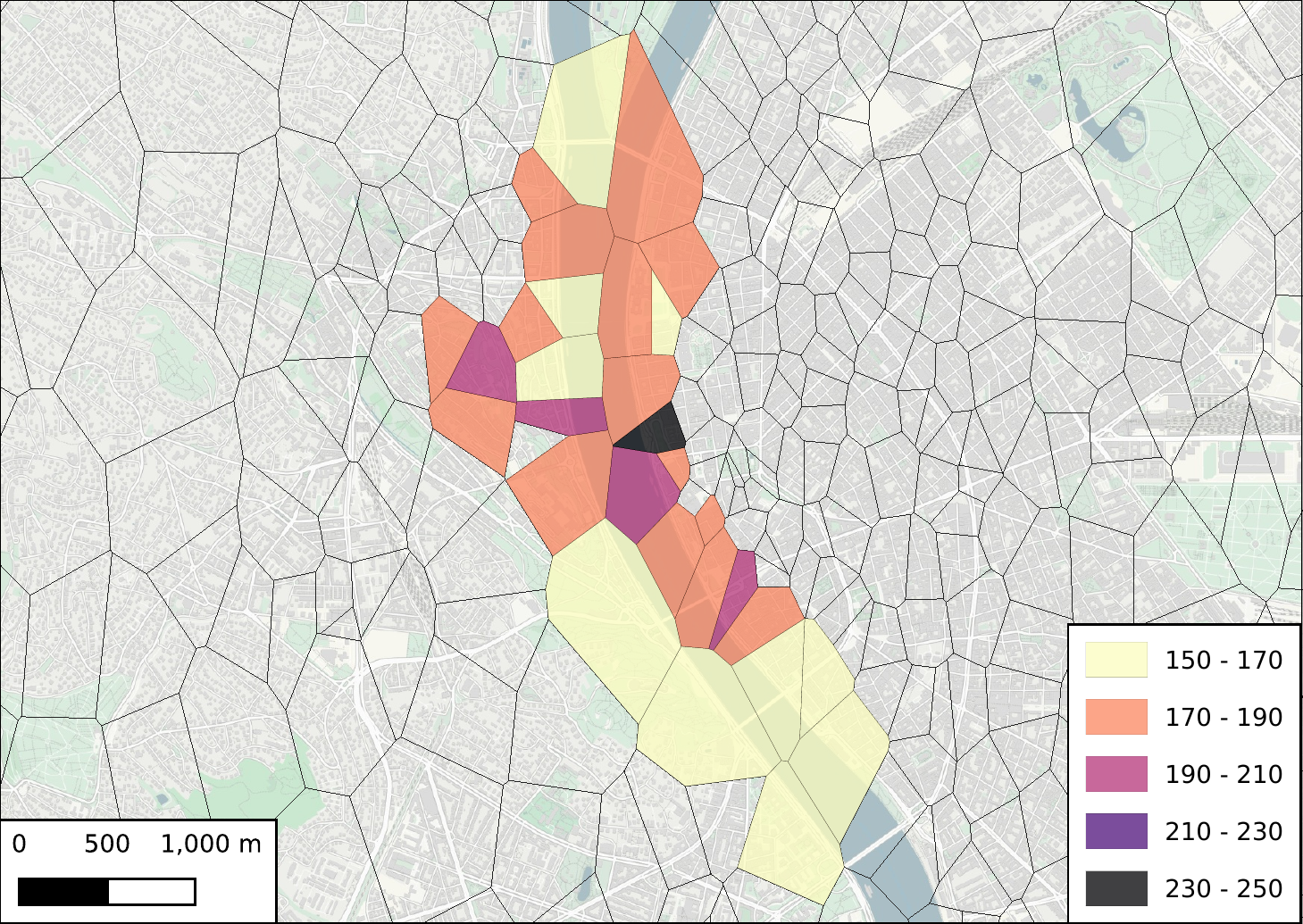}
		\caption{Mobile phone prices, a higher value means higher SES.}
		\label{subfig:firework_price}
	\end{subfigure}
	\hfill
	\begin{subfigure}[t]{0.49\linewidth}
		\centering
		\includegraphics[width=\linewidth]{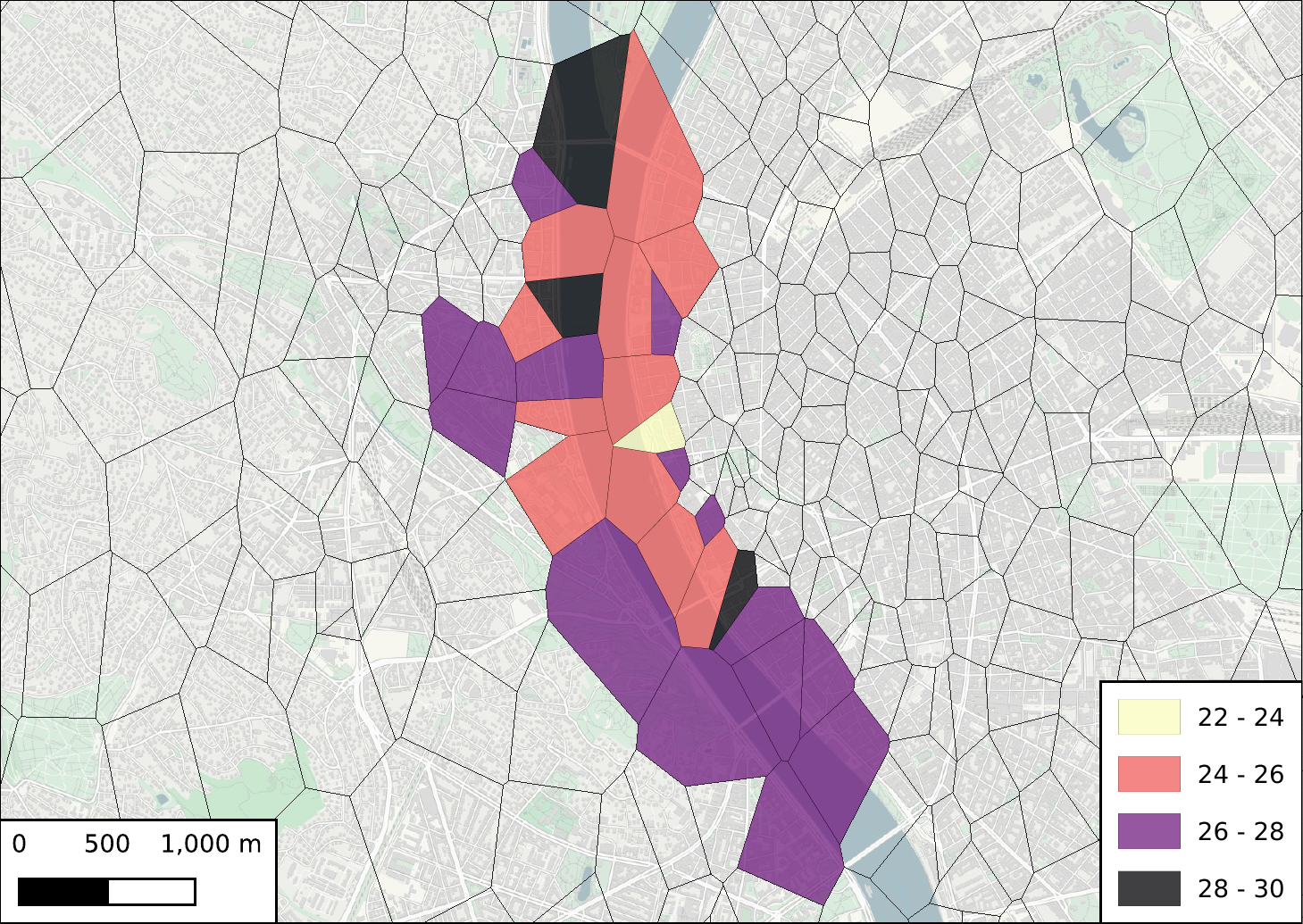}
		\caption{Mobile phone relative ages to the event date (August 2014) in months, a higher value means lower SES.}
		\label{subfig:firework_age}
	\end{subfigure}
	\caption{Average socioeconomic status indicator distributions by riverside cells among the large social event attendees.}
	\label{fig:firework}
\end{figure}

The analysis focused on the socioeconomic status indicator distribution among the State Foundation Day celebratory firework viewers in Budapest, on the banks of the river Danube and in the Castle District in August 2014. The time frame is 20:00 -- 21:30, including half an hour before and after the 30-minute show, marked with vertical lines in Figure \ref{fig:event_day}.

A cell-by-cell average of mobile phone prices and relative ages was calculated for the SES indicator distribution. Figure \ref{fig:firework} shows the spatial distribution of these indicators using Voronoi polygons generated around the cell tower locations. Cells are coloured by the average SES indicators; the higher the value, the darker the colour.

\begin{figure}[H]
	\centering
	\includegraphics[width=0.4\linewidth]{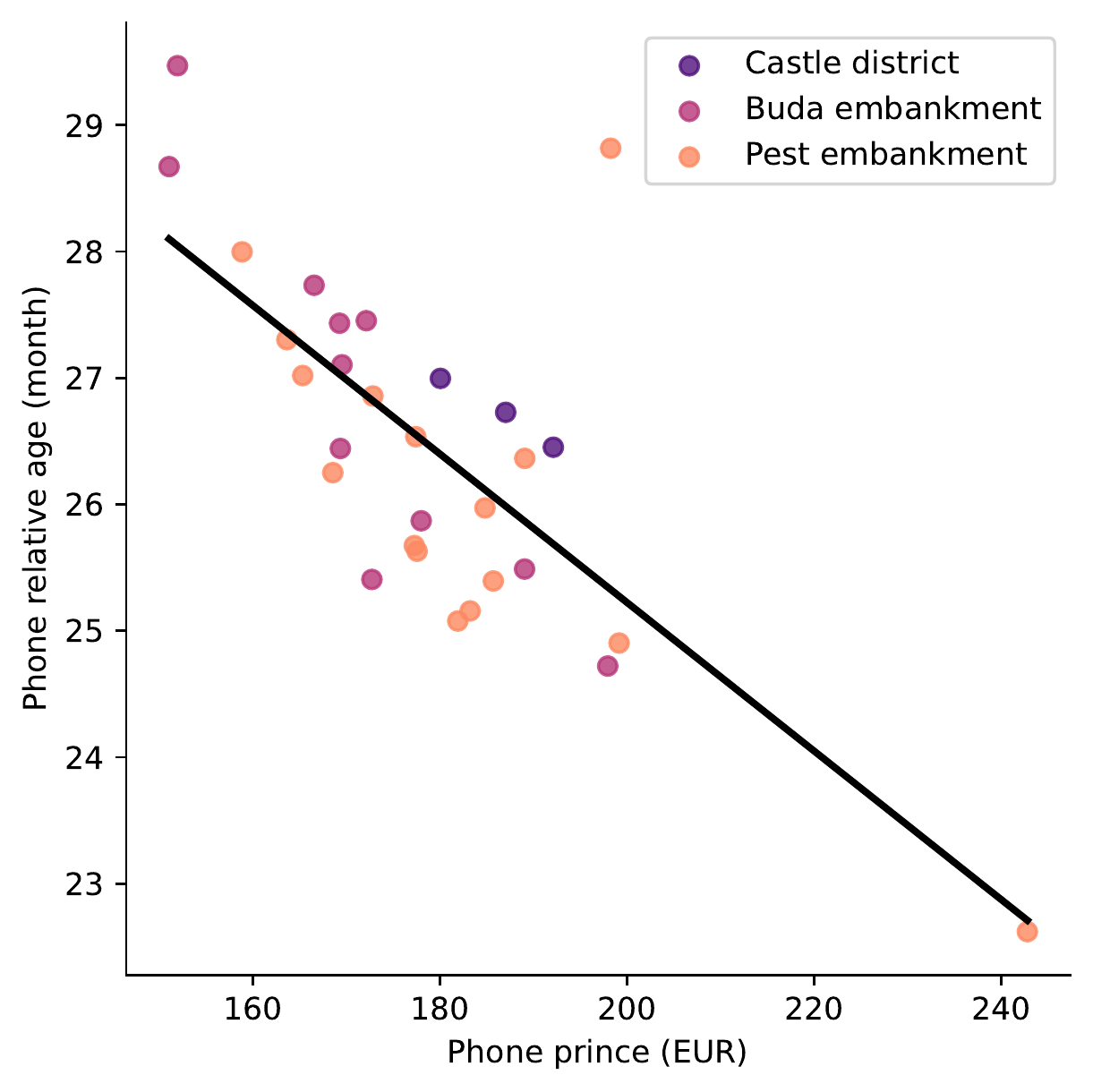}
	\caption{The correlation between average phone prices and ages in cells, where Pearson's $r = -0.7329$.}
	\label{fig:correlation}
\end{figure}

Figure \ref{fig:firework} demonstrates an opposite trend between mobile phone price and age. The scatter plot of the same data is shown in Figure \ref{fig:correlation}, where the Pearson correlation coefficient $= -0.7329$. Data points are coloured based on their location in the city, but there are no visible groups based on SES.

The expectation was that there would be a significant contrast between Buda and Pest in socioeconomic indicator distribution. However, there are only minor differences in this spatial resolution, from which it can be concluded that those interested in the event do not divide drastically into socioeconomic groups.

\section{Conclusions and future work}

This paper presented a concept of socioeconomic data analysis on a large social event using call detail records and mobile phone details. This work fits into current research tendencies, fusing mobile network generated mobility data with socioeconomic descriptors that make it possible to deduce socioeconomic status from anonymous call detail records.

We expected that in Buda, where the housing prices are higher \cite{pinter2021evaluating}, more expensive and newer phones would generate the majority of the activity. Nonetheless, we found that slightly more expensive phones were active in Pest, but the difference, on average, was not substantial. The base station level aggregation may have partly caused this result, or the attendees might not have watched the fireworks isolated from each other based on social status.

For future work, the firework attendees could be grouped into visitors and homeowners in the activity areas. Calculating home positions on CDRs has already been demonstrated to be helpful in \cite{pinter2018evaluation} and could make a difference in the conclusion of this study.

\section*{Author contributions}

Methodology, K.S. and G.P.;
Conceptualisation, G.P. and K.S.;
Software -- data processing, K.S.;
Software -- data preprocessing, G.P.;
Validation, K.S. and G.P.;
Visualisation, K.S.;
Writing, K.S.;
Supervision, G.P. and I.F.

\section*{Acknowledgement}

The authors would like to thank Vodafone Hungary and 51Degrees for providing the Call Detail Records and the Type Allocation Code database used in this study. Map tiles by CartoDB, under CC BY 3.0.

\printbibliography

@article{gonzalez2008understanding,
	title={Understanding individual human mobility patterns},
	author={Gonzalez, Marta C and Hidalgo, Cesar A and Barabasi, Albert-Laszlo},
	journal={nature},
	volume={453},
	number={7196},
	pages={779--782},
	year={2008},
	publisher={Nature publishing group}
}

@article{candia2008uncovering,
	title={Uncovering individual and collective human dynamics from mobile phone records},
	author={Candia, Julián and González, Marta C and Wang, Pu and Schoenharl, Timothy and Madey, Greg and Barabási, Albert-László},
	journal={Journal of physics A: mathematical and theoretical},
	volume={41},
	number={22},
	pages={224015},
	year={2008},
	publisher={IOP Publishing}
}

@inproceedings{pinter2018evaluation,
	title={Evaluation of mobile phone signals in urban environment during a large social event},
	author={Pintér, Gergő and Nadai, László and Bognár, Gábor and Felde, Imre},
	booktitle={2018 IEEE 12th International Symposium on Applied Computational Intelligence and Informatics (SACI)},
	pages={247--250},
	year={2018},
	organization={IEEE}
}

@article{pinter2021evaluating,
	title={Evaluating the Effect of the Financial Status to the Mobility Customs},
	author={Pintér, Gergő and Felde, Imre},
	journal={ISPRS International Journal of Geo-Information},
	volume={10},
	number={5},
	pages={328},
	year={2021},
	publisher={Multidisciplinary Digital Publishing Institute}
}

@article{pinter2021analyzing,
	title={Analyzing the Behavior and Financial Status of Soccer Fans from a Mobile Phone Network Perspective: Euro 2016, a Case Study},
	author={Pintér, Gergő and Felde, Imre},
	journal={Information},
	volume={12},
	number={11},
	pages={468},
	year={2021},
	publisher={Multidisciplinary Digital Publishing Institute}
}

@article{pinter2022awakening,
	title={Awakening City: Traces of the Circadian Rhythm within the Mobile Phone Network Data},
	author={Pintér, Gergő and Felde, Imre},
	journal={Information},
	volume={13},
	number={3},
	pages={114},
	year={2022},
	publisher={MDPI}
}

@inproceedings{sultan2015mobile,
	title={Mobile phone price as a proxy for socio-economic indicators},
	author={Sultan, Syed Fahad and Humayun, Hamza and Nadeem, Umar and Bhatti, Zubair Khurshid and Khan, Sohaib},
	booktitle={Proceedings of the Seventh International Conference on Information and Communication Technologies and Development},
	pages={1--4},
	year={2015}
}

@online{gsmarena,
	author = {Mohit Sainani},
	title = {GSMArena Mobile Phone Devices},
	date = {2020},
	url = {https://www.kaggle.com/msainani/gsmarena-mobile-devices},
	note = {Accessed on 02.03.2022}
}

@article{barnett2016social,
	title={Social and spatial clustering of people at humanity’s largest gathering},
	author={Barnett, Ian and Khanna, Tarun and Onnela, Jukka-Pekka},
	journal={PLOS ONE},
	volume={11},
	number={6},
	pages={1--12},
	year={2016},
	publisher={Public Library of Science San Francisco, CA USA}
}

@article{mamei2016estimating,
	title={Estimating attendance from cellular network data},
	author={Mamei, Marco and Colonna, Massimo},
	journal={International Journal of Geographical Information Science},
	volume={30},
	number={7},
	pages={1281--1301},
	year={2016},
	publisher={Taylor \& Francis}
}

@article{wirz2013probing,
	title={Probing crowd density through smartphones in city-scale mass gatherings},
	author={Wirz, Martin and Franke, Tobias and Roggen, Daniel and Mitleton-Kelly, Eve and Lukowicz, Paul and Tröster, Gerhard},
	journal={EPJ Data Science},
	volume={2},
	number={1},
	pages={1--24},
	year={2013},
	publisher={SpringerOpen}
}

@inproceedings{hiir2019impact,
	title={Impact of Natural and Social Events on Mobile Call Data Records -- An Estonian Case Study},
	author={Hiir, Hendrik and Sharma, Rajesh and Aasa, Anto and Saluveer, Erki},
	booktitle={Complex Networks and Their Applications VIII},
	pages={415--426},
	year={2019},
	organization={Springer}
}

\end{document}